\documentclass[aps,prl,reprint,groupedaddress]{revtex4-1}

\usepackage{graphicx}
\usepackage{amsmath}

\begin{document}

\title{Polarization effects on the effective temperature of an ultracold electron source}

\author{W.J. Engelen}
\author{D.J. Bakker}
\author{O.J. Luiten}
\author{E.J.D. Vredenbregt}
\email[]{e.j.d.vredenbregt@tue.nl}
\affiliation{Department of Applied Physics, Eindhoven University of Technology, P.O. Box 513, 5600 MB Eindhoven, The Netherlands}

\date{\today}

\begin{abstract}
The influence has been studied of the ionization laser polarization on the effective temperature of an ultracold electron source, which is based on near-threshold photoionization. This source is capable of producing both high-intensity and high-coherence electron pulses, with applications in for example electron diffraction experiments. For both nanosecond and femtosecond photoionization, a sinusoidal dependence of the temperature on polarization angle has been found. For most experimental conditions, the temperature is minimal when the polarization coincides with the direction of acceleration. However, surprisingly, for nanosecond ionization a regime exists when the temperature is minimal when the polarization is perpendicular to the acceleration direction. This shows that in order to create electron bunches with the highest transverse coherence length, it is important to control the polarization of the ionization laser. The general trends and magnitudes of the temperature measurements are described by a model, based on the analysis of classical electron trajectories; this model further deepens our understanding of the internal mechanisms during the photoionization process.
Furthermore, for nanosecond ionization, charge oscillations as a function of laser polarization have been observed; for most situations the oscillation amplitude is small.
\end{abstract}

\maketitle

\section{Introduction}

The development of ultrafast electron and X-ray sources allows the investigation of processes that occur at the ultrafast and ultra-small scales (sub-picosecond and nanometre) \cite{Lorenz_PNAS_13, Hanisch-Blicharski_U_13, Spence_RPP_12, Erasmus_PRL_12}. Additionally, there are many important and interesting biological processes, chemical reactions and materials dynamics that occur at the nanosecond scale \cite{Kim_S_08}. With a Dynamic Transmission Electron Microscope (DTEM) \cite{King_JAP_05}, these processes can currently be investigated with in the order of 10 nanosecond temporal resolution and 10 nanometre spatial resolution \cite{LaGrange_U_08}. 

The ultracold electron source would be very well suited to investigate (ultra)fast processes with electron diffraction, as it allows the production of high-intensity, high-coherence electron bunches, by means of near-threshold photoionization of laser-cooled atoms \cite{Claessens_PRL_05}. 
In order to produce high-quality diffraction patterns, the transverse coherence length $L_{\perp}$ of the electron bunch should be larger than the lattice constant of the crystal under investigation. In turn, $L_{\perp}$ can be expressed in terms of the effective source temperature $T$, which is then a measure for the quality of the source; this gives $L_{\perp} = \hbar \sigma_x / \sigma_{x_0} \sqrt{m k_{\rm{B}} T}$, with $\sigma_x$ the root-mean-square (rms) size of the electron beam at the sample, $\sigma_{x_0}$ the size of the beam at the source, $m$ the electron rest mass, and $k_{\rm{B}}$ Boltzmann's constant \cite{Engelen_NC_13}. 

Previously, the source temperature of the ultracold source has been measured with nanosecond photoionization to be as low as $T = 10$\,K \cite{Taban_EPL_10, McCulloch_NP_11, Saliba_OE_12, Engelen_U_14}, while for femtosecond ionization $T = 30$\,K can be achieved \cite{Engelen_NC_13, McCulloch_NC_13}. In contrast, photocathode sources and DTEMs have a source temperature in excess of 1250\,K \cite{Carbone_CP_12}.
This shows the advantage of the ultracold source, both for nanosecond and sub-picosecond electron pulses, as the lower source temperature leads to larger coherence lengths. This allows the investigation of samples with larger lattice constants or improves the diffraction data quality. Alternatively, for a fixed source temperature one could reduce the size of the electron beam at the sample position, when a smaller coherence length is sufficient. As preferably every electron in the pulse is used for diffraction, a smaller beam size allows smaller samples; for example for biological systems, samples are hard to make with a large size.

In this work, the influence of the ionization laser polarization on the source temperature is studied. For both nanosecond and femtosecond photoionization, the temperature has been measured as a function of the polarization angle for a range of wavelengths of the ionization laser. We find that the temperature oscillates sinusoidally with the polarization angle. A classical model is used to describe the general trends and magnitudes of the data. This work not only helps us to optimize the performance of the ultracold electron source, but also provides more insight in the physics that plays a role in near-threshold photoionization; both of which are essential for further development of this source. Furthermore, the influence of the polarization on the bunch charge is discussed.

\section{Experimental setup}

\begin{figure*}
	\centering
	\includegraphics{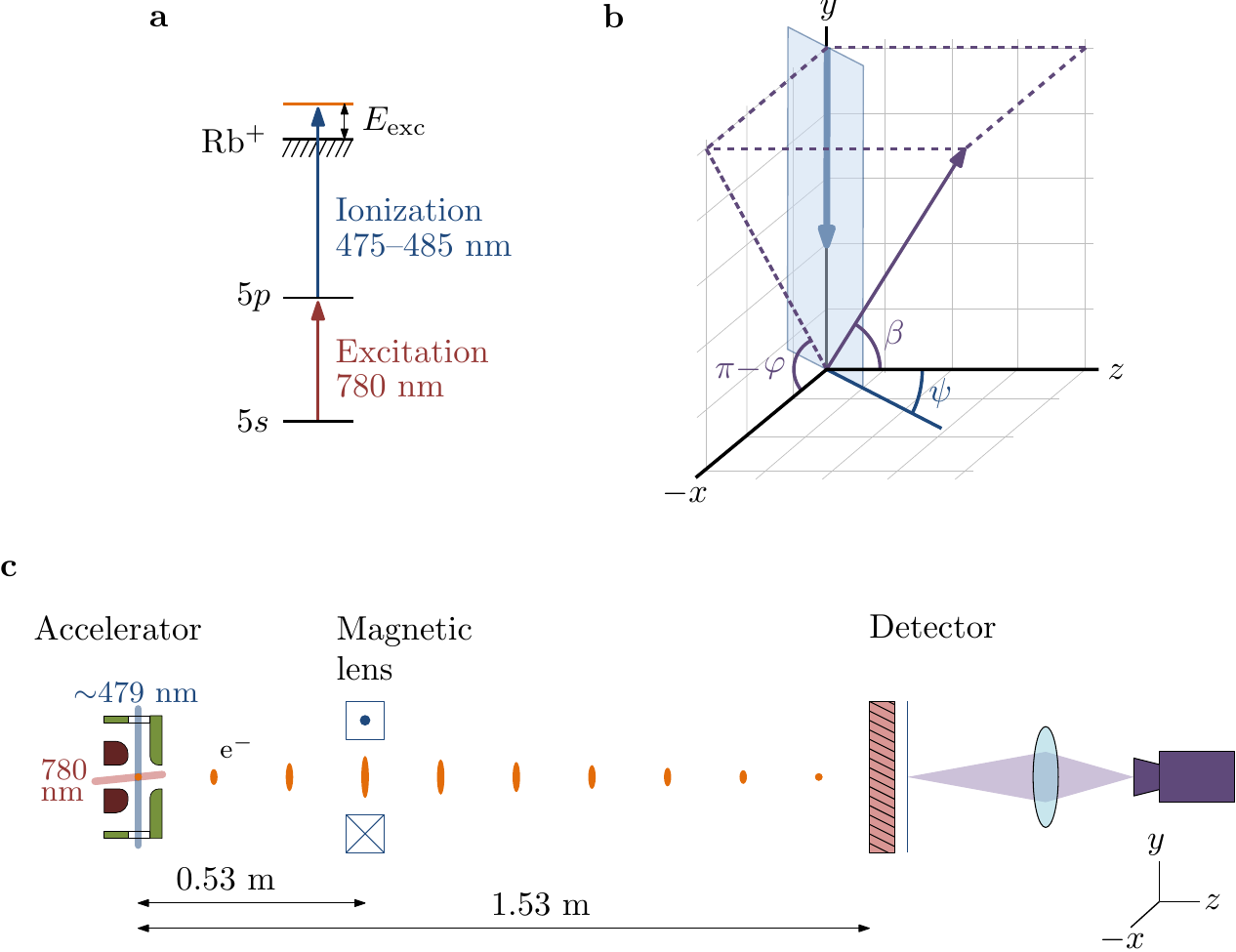}%
	\caption{Experimental setup. a) Ionization scheme. Atoms are excited to the 5$p$ state, and ionized by an ionization laser pulse. The electrons gain an excess energy $E_{\rm{exc}}$ in the ionization process. b) Coordinate system. The ionization laser (blue line) travels in the negative $y$-direction; its polarization is making an angle $\psi$ with the $z$-axis. For the analytical model, the electron's initial velocity is shown (purple line), which is making an angle $\beta$ with the $z$-axis and an angle $\varphi$ with the $x$-axis when projected on the $x$--$y$-plane. c) Electron beam line. Electrons are created in the accelerator and travel to a detector, consisting of a micro-channel plate (MCP) detector with a phosphor screen, which is imaged with a lens on a CCD camera. A magnetic lens is used to move the focal point of the electron beam.\label{fig:setup_pol}}
\end{figure*}

In this section, a brief overview of the experimental setup and the measurement method is given. More details can be found in Ref. \cite{Engelen_U_14}. Electron beams are created from an ultracold source by near-threshold photoionization of laser-cooled atoms. First $^{85}$Rb atoms are laser-cooled and trapped in a vapour-cell magneto-optical trap (MOT), leading to an atom cloud with an rms radius of 1\,mm, containing about $10^8$ particles. Then, a part of the trapped atoms are excited to the 5$p$ state with a 780\,nm diode laser (Fig.\,\ref{fig:setup_pol}a), propagating at a small angle with the $z$-axis (Fig.\,\ref{fig:setup_pol}c), and subsequently ionized close to the ionization threshold by a laser pulse with wavelength $\lambda$ (${\sim}479$\,nm). The ionization laser is linearly polarized in the $x$--$z$-plane, and its polarization angle $\psi$ can be set by a half-wave plate; for $\psi = 0$ the polarization vector points in the $z$-direction (Fig.\,\ref{fig:setup_pol}b). The energy the electrons gain in the photoionization process is characterized by the excess energy $E_{\rm{exc}}$ (see Fig.\,\ref{fig:setup_pol}a); it is the sum of the ionization energy with respect to the ionization threshold $E_\lambda$ in zero electric field and the Stark shift $E_F$ of the ionization threshold by the acceleration field, 
\begin{equation}
	E_{\rm{exc}} = E_\lambda + E_F = h c \left(\frac{1}{\lambda} - \frac{1}{\lambda_0} \right) + 2 E_{\rm{h}} \sqrt{\frac{F_{\rm{acc}}}{F_0}},
\end{equation}
with $\lambda_0$ the ionization threshold wavelength (479.06\,nm for $^{85}$Rb in the 5$p$ state), $E_{\rm{h}} = 27.2$\,eV the Hartree energy, and $F_0 = 5.14 \times 10^{11}$\,V/m the atomic unit of field strength. Finally, the electrons are extracted and accelerated by an electric field $F_{\rm{acc}}$. After leaving the accelerator, the electrons pass a magnetic lens and arrive at $z = 1.53$\,m at the detector, consisting of a micro-channel plate (MCP) detector with a phosphor screen, imaged onto a CCD camera 
(Fig.\,\ref{fig:setup_pol}c).

An upper limit of the effective source temperature can be found by measuring the beam size at the detector as a function of the current through a magnetic lens, a so-called waist scan \cite{Engelen_U_14}. This measurement is comparable to measuring the Rayleigh length of a laser beam, with a larger Rayleigh length corresponding to a higher coherence (lower temperature); here the beam size is not measured at different positions around the waist, but the beam size is measured at a fixed position while varying the lens strength. 

In this work, we present data for nanosecond and femtosecond photoionization. In the former case, atoms are ionized by a pulsed dye laser with a full-width-at-half-maximum (FWHM) pulse length of 5\,ns. The electron beam was extracted from an initial volume with transverse rms dimensions of $27 \pm 2\,\mu{\rm{m}} \times 26 \pm 2\,\mu$m, determined by the overlap of the excitation and ionization laser. The beam had an energy $U = 2.4$\,keV after acceleration with $F_{\rm{acc}} = 0.185$\,MV/m (corresponding $E_F = 32.6$\,meV). For femtosecond photoionization, 58\,fs FWHM laser pulses are used, created with an optical parametric amplifier, pumped by a Ti:Sa laser. In this case, $14 \pm 1\,\mu{\rm{m}} \times 7 \pm 1\,\mu$m electron beams were created with $U = 2.84$\,keV ($F_{\rm{acc}} = 0.222$\,MV/m, $E_F = 35.8$\,meV). The femtosecond laser pulses had a rms bandwidth of 4\,nm, as observed on a spectrometer, leading to electron bunches with a substantial spread in excess energy. This is in contrast to nanosecond ionization, where the spread in excess energy due to the bandwidth of the nanosecond laser (${\sim}$1\,pm or ${\sim}$2\,GHz) can be neglected.

\section{Temperature measurements}

\begin{figure}
	\includegraphics{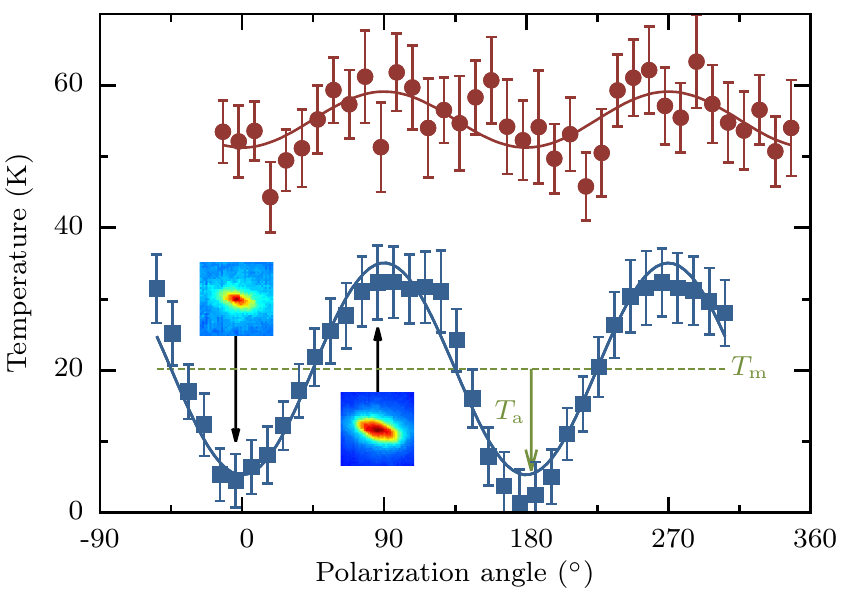}%
	\caption{Temperature as a function of polarization angle $\psi$ for the nanosecond ionization laser for $E_{\rm{exc}} = 6.2$\,meV (blue squares), and for the femtosecond laser for $E_{\rm{exc}} = 25.3$\,meV (red circles). The data are fitted with Eq. \eqref{eq:T_fit} (solid lines), yielding the mean temperature $T_{\rm{m}}$ and the oscillation amplitude $T_{\rm{a}}$. Also shown are two detector images of the electron beam, each corresponding to an extremum in temperature.
\label{fig:T_vs_pol}}
\end{figure}

A measurement of the effective source temperature as a function of the polarization angle of the nanosecond ionization laser for $E_{\rm{exc}} = 6.2$\,meV ($\lambda = 484$\,nm) is shown in Fig. \ref{fig:T_vs_pol} (blue squares). It can be seen that there is a large variation in temperature, with a minimum when the polarization is pointing in the $z$-direction and a maximum when it is pointing in the $x$-direction. The data can be fitted by assuming a simple cosine dependence:
\begin{equation}
	T = T_{\rm{m}} - T_{\rm{a}} \cos{2 \psi},\label{eq:T_fit}
\end{equation}
with $T_{\rm{m}}$ the mean temperature and $T_{\rm{a}}$ the oscillation amplitude. For this case, we find $T_{\rm{m}} = 20 \pm 5$\,K and $T_{\rm{a}} = 15 \pm 2$\,K. The uncertainty indicates the 95\% confidence bound of the fit. In the figure, we have also shown two detector images of the electron beam, each corresponding to an extremum in temperature. These images were measured at a magnetic lens current of 0\,A, in which case the spot size at the detector depends strongly on the source temperature as shown in Fig. (3) of Ref. \cite{Engelen_U_14}. All electron beam images corresponding to the data presented in this paper can be described well with a Gaussian distribution.

\begin{figure}
	\includegraphics{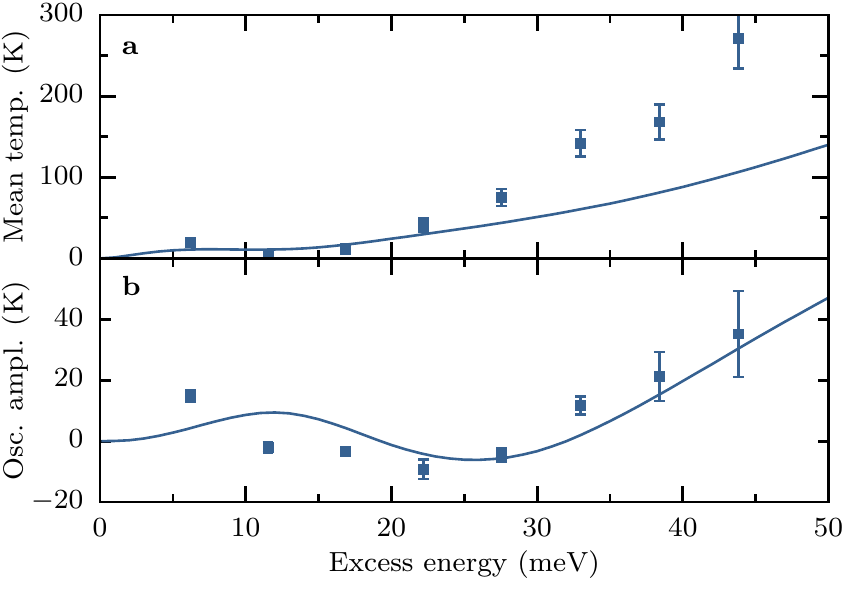}%
	\caption{Mean temperature $T_{\rm{m}}$ (a) and oscillation amplitude $T_{\rm{a}}$ (b) as a function of the excess energy $E_{\rm{exc}}$, measured with the nanosecond ionization laser (points) and calculated with the model (solid lines).\label{fig:ns_T_vs_Eexc}}
\end{figure}

This measurement has been repeated for $E_{\rm{exc}} = 6$--44\,meV ($\lambda = 484$--477\,nm). Each data set shows a sinusoidal temperature oscillation as a function of polarization, and has been fitted with Eq. \eqref{eq:T_fit}. The oscillation amplitude as a function of $E_{\rm{exc}}$ is shown in Fig. \ref{fig:ns_T_vs_Eexc}b (points). For small excess energies, the amplitude is positive. For $E_{\rm{exc}} = 12$--28\,meV ($\lambda = 483$--480\,nm), the amplitude is negative, meaning that the temperature is minimal when the polarization is pointing in the $x$-direction. For larger excess energies, the amplitude is positive again. The values for the mean temperature are shown in Fig. \ref{fig:ns_T_vs_Eexc}a (points). The temperatures found for $\psi = 0^\circ$, that is $T_{\rm{m}} - T_{\rm{a}}$, are in good agreement with previous results \cite{Engelen_U_14}, where this quantity was also determined.

\begin{figure}
	\includegraphics{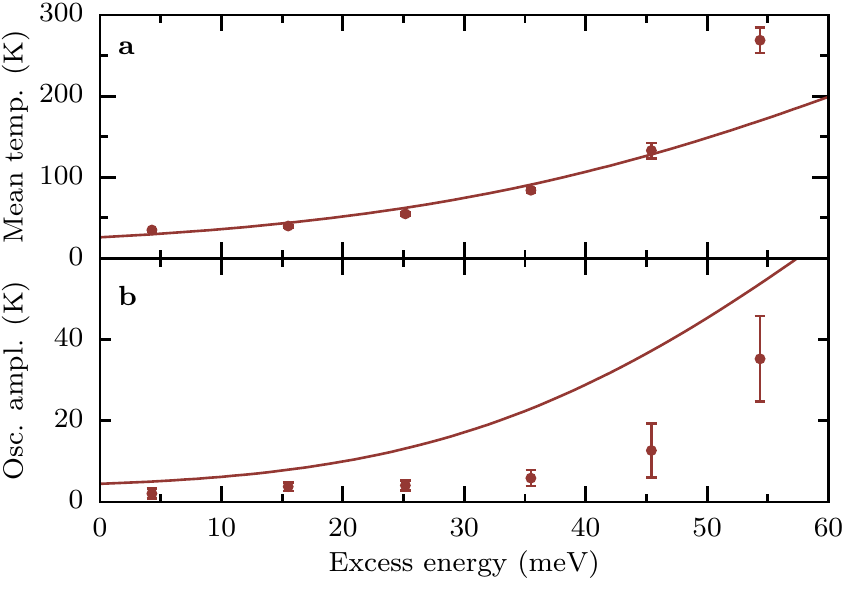}%
	\caption{Mean temperature $T_{\rm{m}}$ (a) and oscillation amplitude $T_{\rm{a}}$ (b) as a function of the excess energy $E_{\rm{exc}}$, measured with the femtosecond ionization laser (points) and calculated with the model (solid lines).\label{fig:fs_T_vs_Eexc}}
\end{figure}

Next, these measurements have been repeated with the femtosecond ionization laser for $E_{\rm{exc}} = 4$--54\,meV ($\lambda = 485$--476\,nm). Again each data set showed a sinusoidal oscillation as a function of polarization. As an example, the data set for $E_{\rm{exc}} = 25.3$\,meV ($\lambda = 481$\,nm) has been plotted in Fig. \ref{fig:T_vs_pol} (red circles). The oscillation amplitude (Fig. \ref{fig:fs_T_vs_Eexc}b, points) is positive for all measured wavelengths. We again find that $T_{\rm{m}} - T_{\rm{a}}$, with $T_{\rm{m}}$ shown in Fig. \ref{fig:fs_T_vs_Eexc}a (points), is comparable to previous 
measurements \cite{Engelen_NC_13}.

\section{Temperature model}

To explain the experimental results, a model has been developed to calculate the temperature of the ultracold electron source. The model is based on classical electron trajectory calculations;
for details, see Ref. \cite{Engelen_bordas_model, Bordas_PRA_98}. The model explains why high coherence bunches can be made by near-threshold photoionization, both for narrowband ionization pulses and for ultrashort, broadband 
pulses \cite{Engelen_bordas_model}. Here we will incorporate the polarization of the ionization laser. 

\begin{figure}
	\includegraphics{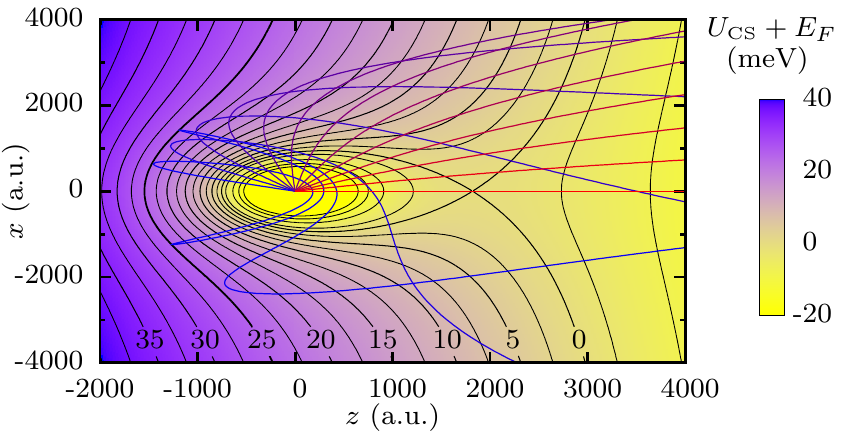}%
	\caption{Potential landscape with equipotential lines for $U_{\rm{CS}} + E_F$, with $E_{\rm{exc}} = 25$\,meV and $F =$ 0.185\,MV m$^{-1}$, together with electron trajectories calculated with the classical model, for different starting angles $0 \le \beta \le \beta_{\rm{c}} = 161^{\circ}$.\label{fig:pol_potential_surface}}
\end{figure}

The dynamics of the electrons just after photoionization is taken into account in this model. By analysing the classical trajectories of electrons that move in the potential $U_{\rm{CS}}$ of the ion and the acceleration field, the asymptotic transverse velocity $v_r$ of electrons can be derived \cite{Engelen_bordas_model}. This velocity depends on the acceleration field, excess energy, and angle $\beta$, which is the starting angle between the initial electron velocity and the direction of acceleration 
(Fig.\,\ref{fig:setup_pol}b). As an example, electron trajectories calculated with the model for $E_{\rm{exc}} = 25$\,meV and $F =$ 0.185\,MV m$^{-1}$ are shown in Fig. \ref{fig:pol_potential_surface}.

When using the model to calculate $v_r$ for electrons escaping from a hydrogen-like atom, the system can be solved analytically. In that case, only electrons with an angle smaller than a critical angle $\beta_{\rm{c}} = 2 \arccos \left( \frac{-E_{\lambda}}{E_F} \right)$ can escape the potential when $E_\lambda \le 0$; the others cannot escape the atom, because they follow closed orbits around the atom, which are allowed by the $1/r$ potential. For $E_\lambda > 0$, all electrons can escape, so $\beta_{\rm{c}} = 180^\circ$. When using the model with a potential appropriate for rubidium, $v_r$ cannot be calculated analytically and has to be determined with simulations. Due to the core electrons, the rubidium potential is no longer strictly a $1/r$ potential. As a result, all electrons with $E_{\rm{exc}} > 0$ can eventually leave the ion. As the resulting temperatures of the model when using the hydrogen and the rubidium potential only differ a few percent at most \cite{Engelen_bordas_model}, the former potential is used in the temperature calculations.

From the asymptotic transverse velocity $v_r$ as a function of $\beta$, the temperature $T = m \sigma_{v_x}^2 / k_{\rm{B}}$ can be obtained, with
\begin{align}
	&\sigma_{v_{x}}^2 = \frac{1}{2} \int_{0}^{\beta_{\rm{c}}} v_r(\beta)^2 W(\beta,\psi) \, {\rm{d}}\beta, \nonumber \\
	&W =  \left\{
					\begin{array}{ll}
						w(\beta,\psi) \sin{\beta} \, \big/ \int_{0}^{\beta_{\rm{c}}} w(\beta',\psi) \sin{\beta'} \, {\rm{d}}\beta' & \text{if } \beta < \beta_{\rm{c}}\\
						0 & \text{if } \beta \ge \beta_{\rm{c}}
\end{array} \right.\label{eq:bordas_W}
\end{align}
Here, the factor 1/2 arises as $\sigma_{v_{x}} = \sigma_{v_{r}} / \sqrt{2}$, $W(\beta,\psi)$ is the normalized initial angular distribution of the electrons, $w(\beta,\psi)$ a weight function,  and the term $\sin{\beta}$ creates an uniformly filled spherical shell of electrons when $w = 1$. In Ref. \cite{Engelen_bordas_model}, a uniformly distributed $\beta$ was used, so $w = w_{\rm{u}} \equiv 1$. However, to model the results presented in this work, we have to take into account that the electrons are preferentially ejected along the laser polarization axis: $w \sim \cos^2 \alpha$ \cite{Bethe_Book_77}, which is valid when the electron orbits in the initial state before ionization (5$p$ in our case) have no preferential orientation. Here $\alpha$ is the angle between the initial electron velocity and the laser polarization, with $\cos \alpha = \cos{\varphi} \, \sin{\beta} \, \sin{\psi} + \cos{\beta} \, \cos{\psi}$ and $\varphi = \arctan v_y/v_x$ the angle the initial electron velocity makes with the $x$-axis when projected in the $x$--$y$-plane (Fig.\,\ref{fig:setup_pol}b).

\begin{figure}
	\includegraphics{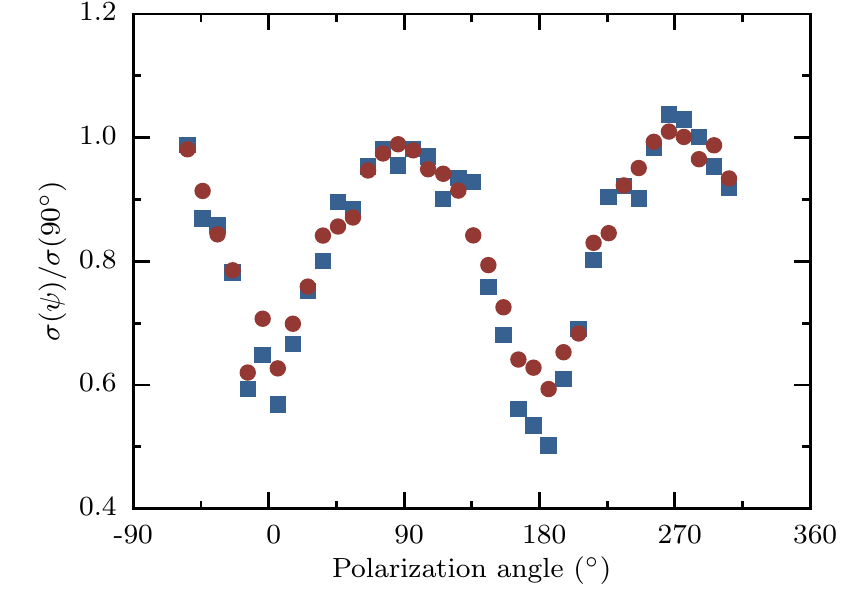}%
	\caption{Measured normalized spot size $\sigma(\psi) / \sigma(90^\circ)$ in the $x$- and $y$-direction (blue squares and red circles) as a function of polarization angle $\psi$ of the nanosecond ionization laser for $E_{\rm{exc}} = 6.2$\,meV.\label{fig:spot_size_vs_polarization}}
\end{figure}

When using the weight function $w \sim \cos^2 \alpha$ in calculating the transverse temperature, a different temperature for the $x$-direction ($T_x$) is found than for the $y$-direction ($T_y$) for $\psi \neq 0$. For example for low excess energies, $T_x$ increases when increasing $\psi$ from 0 to 90$^\circ$, while $T_y$ decreases. This can be understood as electrons are then mostly ejected in the $x$-direction, leading to $\sigma_{v_{x}} > \sigma_{v_{y}}$. However, in the experiments we do not see a difference in temperature between the $x$- and $y$-direction. This is illustrated in Fig. \ref{fig:spot_size_vs_polarization}, where a measurement of the normalized spot size $\sigma(\psi) / \sigma(90^\circ)$ on the detector for the $x$- and $y$-direction is plotted for a magnetic lens current of 0 A, using the nanosecond ionization laser with $E_{\rm{exc}} = 6.2$\,meV. An increase in the spot size corresponds to an increase in temperature, and a decrease in size in a decrease in $T$. It can be seen that the normalized spot sizes both decrease by the same amount when increasing $\psi$ from $90^\circ$ to 180$^\circ$. For the experimental conditions of this measurement, the spot size scales approximately linearly with the source temperature, leading to the conclusion that $T_x = T_y$. Furthermore, the model curve does not have a sinusoidal dependence on polarization angle, whereas this shape has always been observed in experiments.

In, for example, photoionization experiments on metastable Xe, an asymmetry in the $x$--$y$-plane plane has been observed \cite{Lepine_PRA_04}. One major difference in the experimental setup that was used in that work is that the electrons were created in zero magnetic field. In our setup, the trapping coils, which produce the magnetic field needed for the MOT, stay on during the experiment; they produce a quadrupole magnetic field with a gradient of about 10\,G/cm. The magnetic field is approximately zero at the centre of the trapped atom cloud. 
This field could potentially turn the initial asymmetric electron distribution in a symmetric distribution, as there is a magnetic field component in the direction of the electron velocity; this leads to a rotation of the electrons in the $x$--$y$-plane. The strength of this field component is dependent on the distance to the $z$-axis. However, in model simulations of the electron trajectories with a magnetic field present, we have not observed that this process leads to a symmetric distribution, even when the centre of the electron cloud does not coincides with the centre of the quadrupole. The electric field profile in the accelerator, which is slightly inhomogeneous, or the energy spread in the bunch does not provide an explanation for this randomization process either.

Despite the fact that we cannot explain why the initial electron distribution is randomized, we account for our experimental observation of symmetry in the $x$--$y$-plane by averaging the weight function $w = 2 \cos^2 \alpha$ over $\varphi$, leading to
\begin{equation}
	w_{\rm{p}}(\psi) = \sin^2{\beta} \, \sin^2{\psi} + 2 \cos^2{\beta} \, \cos^2{\psi}.\label{eq:weigh_pol}
\end{equation}
When using this weight function in the model, a sinusoidal oscillation of temperature as function of polarization angle is found, with $T_x = T_y$ and extreme values at $\psi = 0^\circ$  and $90^\circ$. To illustrate the origin of the oscillation, we show $v_r(\beta)$ for $F_{\rm{acc}} = 0.185$\,MV/m and $E_{\rm{exc}} = 5$\,meV in Fig. \ref{fig:bordas_model_polarization_Vacc_5_Eexc_5}a. In general, $v_r$ increases as $\beta$ increases. For $\psi = 0^\circ$, $w_{\rm{p}}(0^\circ) = 2 \cos^2{\beta}$, and electrons are preferentially ejected under small angles, thus with small transverse velocities. This leads to a lower $T$ compared to $\psi = 90^\circ$, where more electrons are ejected under higher $\beta$, as $w_{\rm{p}}(90^\circ) = \sin^2{\beta}$, and thus with higher $v_r$. To determine $T$, $W$ needs to be calculated; this is shown in Fig. \ref{fig:bordas_model_polarization_Vacc_5_Eexc_5}b, for three cases: 

\begin{figure}
	\includegraphics{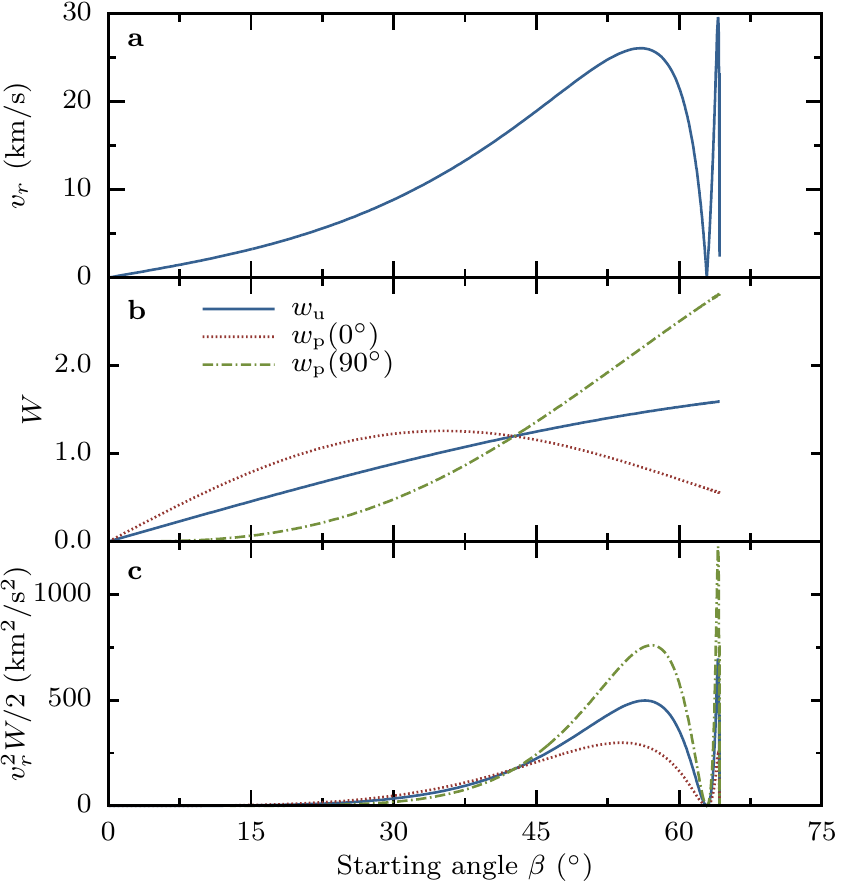}%
	\caption{Model calculation to determine the electron temperature for $F_{\rm{acc}} = 0.185$\,MV/m and $E_{\rm{exc}} = 5$\,meV. a) Transverse electron velocity $v_r$ as a function of starting angle $\beta$. b) W (Eq. \eqref{eq:bordas_W}) for $w = w_{\rm{u}}$, $w = w_{\rm{p}}(0^\circ)$, and $w = w_{\rm{p}}(90^\circ)$. c) $v_r^2 W / 2$ for the 3 cases used in b).\label{fig:bordas_model_polarization_Vacc_5_Eexc_5}}
\end{figure}

\noindent I)\phantom{II} $w = w_{\rm{u}}$ (solid line) 

\noindent II)\phantom{I} $w = w_{\rm{p}}(0^\circ)$ (dotted line) 

\noindent III) $w = w_{\rm{p}}(90^\circ)$ (dashed-dotted line). 

\noindent With these values for $W$, $v_r^2 W$ has been calculated (Fig. \ref{fig:bordas_model_polarization_Vacc_5_Eexc_5}c), from which the temperature is obtained by integrating over $\beta$.
For case II, more electrons are ejected at smaller starting angles in comparison to case I, and fewer electrons at a larger ($> 45^\circ$) angle. As for this particular value of the excess energy low angles correspond to low transverse velocities and large angles to large velocities, this leads to an electron temperature that is lower for case II (7.3\,K) than for case I (9.8\,K). For case III, more electrons are ejected under a large angle compared to case II, leading to an even higher temperature of 12.9\,K. When rotating the polarization, a temperature oscillation is found with a positive amplitude, given by Eq. \eqref{eq:T_fit} with $T_{\rm{m}} = 10.1$\,K and $T_{\rm{a}} = 2.8$\,K.

\begin{figure}
	\includegraphics{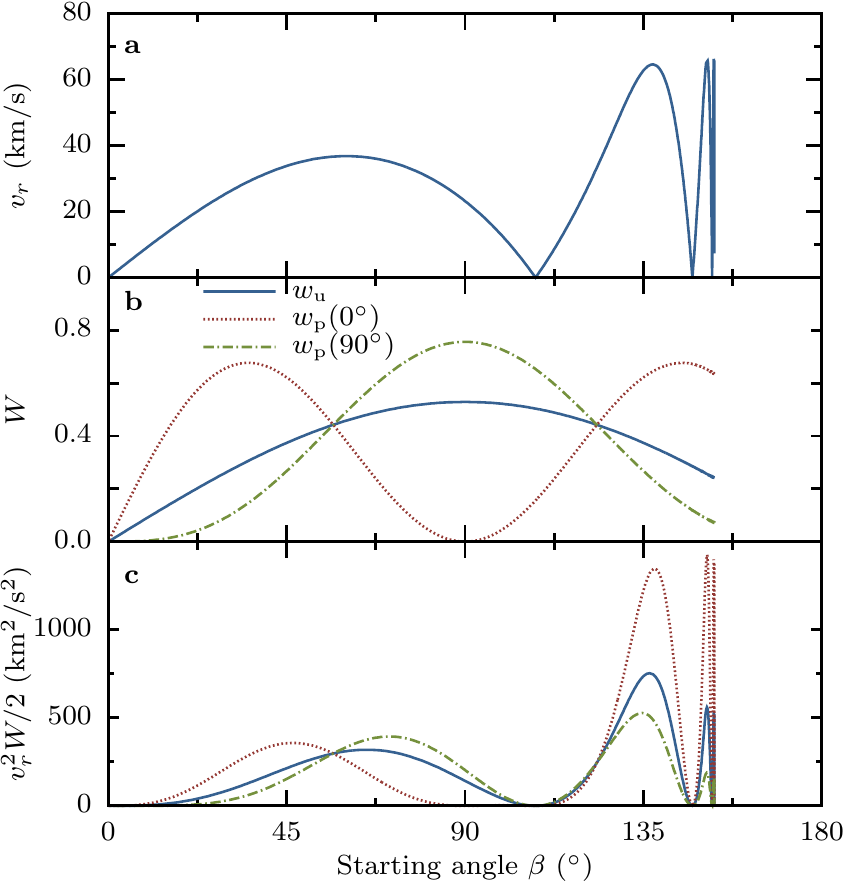}%
	\caption{Same as Fig. \ref{fig:bordas_model_polarization_Vacc_5_Eexc_5}, with calculations performed for $E_{\rm{exc}} = 25$\,meV.\label{fig:bordas_model_polarization_Vacc_5_Eexc_25}}
\end{figure}

In Fig. \ref{fig:bordas_model_polarization_Vacc_5_Eexc_25}, these calculations have been done for $E_{\rm{exc}} = 25$\,meV. This leads to a much larger critical angle $\beta_{\rm{c}}$ compared to $E_{\rm{exc}} = 5$\,meV. Also, the transverse velocity distribution has a different shape, with additional zero crossings.
For case I, we now find $T = 34.6$\,K. Interestingly, for case III, a temperature $T = 30.9$\,K is obtained, which is \emph{lower} than $T = 43.1$\,K for case II. This is caused by the fact that fewer electrons are ejected under angles around 145$^\circ$, which correspond to large transverse velocities. For this situation a negative temperature amplitude is found, with $T_{\rm{m}} = 37.0$\,K and $T_{\rm{a}} = -6.1$\,K.
For high excess energies ($E_{\rm{exc}} > 2 E_F$), the transverse velocity is approximately given by $v_r = \sqrt{2 E_\lambda / m} \sin \beta$. Thus $v_r$ is large for $\beta = 90^\circ$ and small when $\beta = 0^\circ$ or $180^\circ$; combining this with the weight function $w_{\rm{p}}$ one again finds a positive oscillation amplitude.

With the model, the mean temperature and oscillation amplitude have been calculated as a function of excess energy for $F_{\rm{acc}} = 0.185$\,MV/m and nanosecond photoionization (Fig. \ref{fig:ns_T_vs_Eexc}, solid lines). As with previous results \cite{Engelen_U_14}, the model mean temperature follows the same trend as the experimental data. For low excess energies the data lie on the model curve, but for higher excess energies the measured temperatures are higher. The second data point has a lower value than the first; for higher excess energies the temperature increases again with energy. This trend has been observed before \cite{Engelen_U_14} and is also visible in the model: the temperature first increases with energy, then decreases slightly, before increasing again. The model oscillation amplitude has the same trend and magnitude as the data. It shows, as the data does, an amplitude that is first positive, then negative, and then back to positive again, although the model curve seems to be shifted to slightly higher excess energies. The classical, hydrogen-like model employed here therefore already captures the main features of the experimental data. Numerical calculations employing the rubidium model of Ref. \cite{Engelen_bordas_model} extended with the weight function of Eq. \ref{eq:weigh_pol} give essentially the same result. 

To compare the model with the femtosecond photoionization data, the model temperature curve has been convoluted with a Gaussian distribution corresponding to the photon energy spread in the laser pulse \cite{Engelen_bordas_model}, leading to the solid curves in Fig. \ref{fig:fs_T_vs_Eexc}. The model curve for the mean temperature describes the data very well, although for $E_{\rm{exc}} > 50$\,meV the data points lie above the model curve. The model curve for the oscillation amplitude follows the same trend as the data, i.e, predicts an amplitude that is always positive as the data does, but the model values are systematically higher than the data.

One assumption in the model was that the electron orbits in the initial state before ionization (5$p$ in our case) have no preferential orientation. To create electrons in the experiment, the trapping laser, used for the laser cooling, is blocked for 5\,$\mu$s, after which nearly all the atoms are in the ground state. A part of the cloud of atoms is excited by an excitation laser. This laser is circularly polarized, which could potentially optically pump the atoms to the 5$p$, $m = +4$ state, assuming the light is right-circularly polarized. The magnetic field that is present would however counteract this optical pumping to a specific state. We therefore assume that the initial state has no preferential orientation.
To check this assumption experimentally, temperature measurements have been performed with the trapping laser always on. In this situation, the trapping laser populates the $5p$ state, instead of the excitation laser. As three of the six trapping beams are right-circularly polarized and the others left-circularly polarized, and the trapping laser beams are mutually orthogonal, we expect an isotropic distribution over the magnetic quantum numbers. After the ionization laser pulse, a cylinder of electrons is created, with a height equal to the size of the atom cloud, and a width and depth determined by the size of the ionization laser beam. The temperature of the bunch can again be determined by measuring the width of the electron bunch on the detector. The same temperature and charge oscillations (discussed in the next Section) are found as when using the excitation--ionization laser scheme. Therefore we conclude that the initial state in the excitation--ionization laser scheme indeed has no preferential orientation.

\section{Charge measurements}

\begin{figure}
	\includegraphics{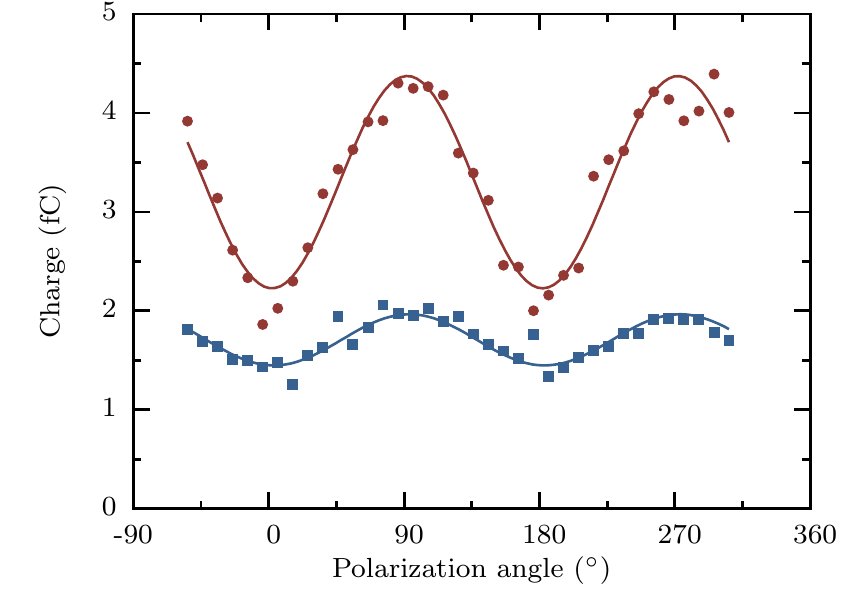}%
	\caption{Bunch charge $Q$ as a function of polarization angle $\psi$ of the nanosecond ionization laser for $E_{\rm{exc}} = 19.5$\,meV (blue squares) and $E_{\rm{exc}} = 6.2$\,meV (red circles), together with a fit using Eq. \eqref{eq:Q_fit} (solid lines).\label{fig:charge_vs_pol}}
\end{figure}

The charge $Q$ of the electron bunch can be measured with a Faraday cup (a metal plate that can be rotated into the beam line) that is connected to a charge amplifier. A measurement of the charge as a function of polarization angle for the nanosecond ionization laser for $E_{\rm{exc}} = 19.5$\,meV is shown in Fig. \ref{fig:charge_vs_pol} (blue squares). An oscillation in charge can be seen, with a minimum when the polarization is in the $z$-direction. The data is fitted (solid line), analogously to Eq. \eqref{eq:T_fit}, with 
\begin{equation}
	Q = Q_{\rm{m}} - Q_{\rm{a}} \cos{2 \psi},\label{eq:Q_fit}
\end{equation}
with $Q_{\rm{m}}$ the mean charge and $Q_{\rm{a}}$ the oscillation amplitude. 

\begin{figure}
	\includegraphics{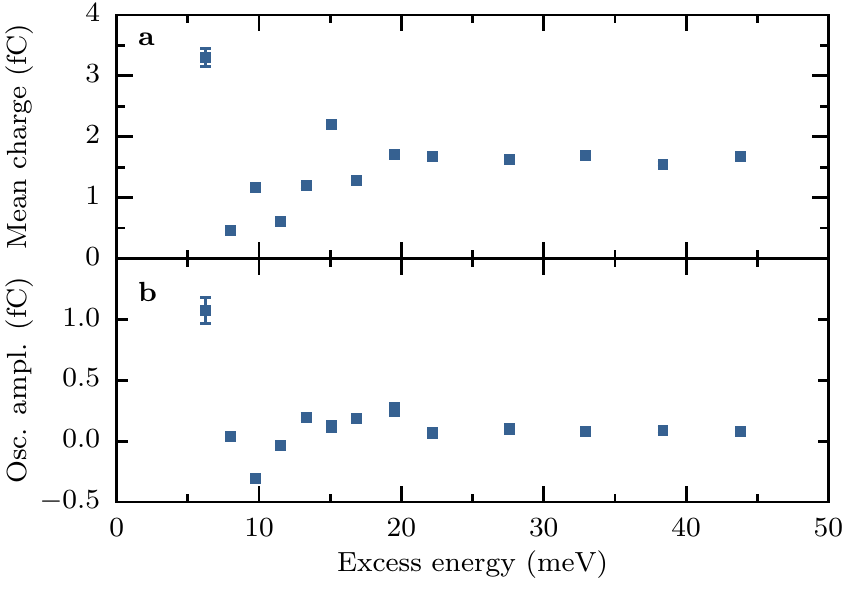}%
	\caption{Mean charge $Q_{\rm{m}}$ (a) and oscillation amplitude $Q_{\rm{a}}$ (b) as a function of the excess energy $E_{\rm{exc}}$, measured with the nanosecond ionization laser.\label{fig:charge_equi_amp}}
\end{figure}

Such measurement have been done for $E_{\rm{exc}} = 6.2$--43.8\,meV. In Fig. \ref{fig:charge_equi_amp}, $Q_{\rm{m}}$ and $Q_{\rm{a}}$ are shown as a function of excess energy. In general, the figure shows that the amount of charge is largely independent of polarization angle, except for excess energies smaller than $\sim 10$\,meV, i.e., very close to threshold. 
For excess energies larger than $15$\,meV, the mean charge is essentially constant and the amplitude small and positive. In this regime, the oscillation can be neglected with respect to the mean value. In the model that has been used to explain the temperature oscillations, we assumed that the initial electron state has no preferential orientation; this leads to a constant charge as function of the polarization. The residual small oscillation could be explained by a polarization dependent overlap between the initial and the final state of the electron. For $E_{\rm{exc}} \sim 10$\,meV, the amplitude becomes negative. Close to the ionization threshold, the mean charge increases and the oscillation amplitude is large and positive. This is shown in Fig. \ref{fig:charge_vs_pol} (red circles) for $E_{\rm{exc}} = 6.2$\,meV. An intuitive explanation is that close to the threshold, there are resonances present in the continuum of states, as the density of states still has some structure, which alter the ionization probability and thus the charge that is produced. To describe this, a full quantum-mechanical treatment is needed. Note that the excess energies for which the temperature and the charge amplitudes change sign are not the same. For the femtosecond ionization laser, no charge oscillations were observed. This is explained by the $22$\,meV rms bandwidth of the photon energy in this case, which smoothes out any variations on this scale.

It has been checked that the temperature for a fixed polarization angle is independent of the charge that is produced, where the charge has been varied by changing the energy of the ionization laser pulse. An oscillation in charge only therefore does not lead to an oscillation in temperature. Additionally, the charge that is produced for a fixed polarization angle is mostly independent of the bunch temperature, as can be seen in Fig. \ref{fig:charge_equi_amp}a for $E_{\rm{exc}} > 20$\,meV.

\section{Conclusion}

We have shown that the temperature of electron bunches produced by near-threshold photoionization varies sinusoidally as the polarization of the ionization laser is rotated, both for nanosecond and femtosecond photoionization. This shows that, in order to create electron bunches with the lowest possible temperature, it is important to take this polarization effect into account. For nanosecond ionization the oscillation amplitude changes from positive to negative as the excess energy is increased, before becoming positive again, after which it increases in value with $E_{\rm{exc}}$. For femtosecond ionization the amplitude is positive and increases in size with $E_{\rm{exc}}$. 

A model has been developed, based on classical trajectory calculation of electrons in the ionization potential. This model predicts a sinusoidal temperature oscillation as a function of polarization angle, and further deepens our understanding of the internal mechanisms during the photoionization process. The general trend and magnitude of the mean temperature and the oscillation amplitude calculated with the model agree with the data; however the nanosecond model amplitude seems to be shifted to slightly higher excess energies with respect to the data and the femtosecond model amplitude is systematically higher than the data.

For nanosecond ionization, charge oscillations have been observed. For larger excess energies, the amplitude is small and positive; for decreasing excess energies, the amplitude becomes negative before becoming large and positive close to the ionization threshold.

This research is supported by the Dutch Technology Foundation STW, which is part of the Netherlands Organisation for Scientific Research (NWO), and which is partly funded by the Ministry of Economic Affairs.


\end{document}